\documentclass[a4paper,11pt]{article}
\usepackage{pos}
\usepackage{slashed}

\usepackage{cancel}

\newcommand{\mc}{\mathfrak{m}}
\newcommand{\ic}{{\rm i}}
\renewcommand{\logo}{\relax}

\title{The limits of the strong $CP$ problem}

\author[a]{Wen-Yuan Ai}
\author[b]{Juan S. Cruz}
\author[c]{Bj\"orn Garbrecht}
\author*[d]{Carlos Tamarit}

\affiliation[a]{Theoretical Particle Physics and Cosmology, King’s College London,\\
Strand, London WC2R 2LS, UK}

\affiliation[b]{CP$^3$-Origins, Center for Cosmology and Particle Physics Phenomenology,\\
University of Southern Denmark,
Campusvej 55, 5230 Odense M, Denmark}

\affiliation[c,d]{Physik-Department T70, Technische Universit\"at M\"unchen,\\ James-Franck-Stra{\ss}e, 85748 Garching, Germany}

\emailAdd{wenyuan.ai@kcl.ac.uk}
\emailAdd{jcr@sdu.dk}
\emailAdd{garbrecht@tum.de}
\emailAdd{carlos.tamarit@tum.de}

\abstract{While $CP$ violation has never been observed in the strong interactions, the QCD Lagrangian admits a $CP$-odd topological interaction proportional to the so called $\theta$ angle, which weighs the contributions to the partition function from different topological sectors. The observational bounds are usually interpreted as demanding a severe tuning of $\theta$ against the phases of the quark masses, which constitutes the strong $CP$ problem. Here we report on recent challenges to this view based on a careful treatment of boundary conditions in the path integral and of the limit of infinite spacetime volume, which leads to $\theta$ dropping out of fermion correlation functions and becoming unobservable, implying that $CP$ is preserved in QCD.}

\FullConference{%
  7th Symposium on Prospects in the Physics of Discrete Symmetries (DISCRETE 2020-2021)\\
  29th November - 3rd December 2021\\
 Bergen, Norway}

\begin{document}

\begin{flushleft}                          
\footnotesize   KCL-PH-TH/2022-31,TUM-HEP-1403-22
\end{flushleft} 
\maketitle

\section{Introduction}
The  Lagrangian of quantum chromodynamics (QCD) with gauge fields $A_\mu$ in the Lie Algebra of SU(3) and $N_f$ flavours of Dirac fermions $\psi_i$ admits  $CP$-noninvariant terms. After a rotation to Euclidean spacetime, used throughout the rest of the paper, these terms read 
\begin{align}
\label{eq:topologcicalterm}
{\cal L}\supset\sum_{i=1}^{N_f}\bar\psi_i\left(\mc_i P_{\rm R}+\mc_i^*P_{\rm L}\right)\psi_i-\frac{\ic}{16\pi^2}\, \theta\, {\rm tr}\,F \widetilde F.
\end{align}
Above, $F_{\mu\nu}=\partial_\mu A_\nu-\partial_\nu A_\mu-{\rm i} [A_\mu,A_\nu]$ is the non-Abelian field strength, 
while $\tilde F_{\mu\nu}=\frac{1}{2}\,{\epsilon_{\mu\nu\rho\sigma}}F_{\rho\sigma}$. 
$P_{\rm R/L}$ are right/left handed projectors, and $\mc_i\equiv m_i {\rm e}^{{\rm i}\alpha_i}, m_i,\alpha_i\in\mathbb{R}$ are complex fermion masses.
$CP$-violating physical observables may be sensitive to combinations of the $CP$-odd parameters $\alpha_n,\theta$ that remain invariant under chiral field redefinitions. As follows from chiral anomalies, the relevant combination is
\begin{align}\label{eq:bartheta}
 \bar\theta=\theta+\sum_i\alpha_i\equiv \theta+\bar\alpha.
\end{align}
 The  $\theta$-dependent contribution  in Eq.~\eqref{eq:topologcicalterm} can be seen to be a total derivative. As such, it is a boundary term whose effects can only be captured with nonperturbative computations. Moreover, whenever the fields go to pure gauge configurations at the boundary $\partial\Omega$ of the four-dimensional spacetime $\Omega$, the $\theta$-term is proportional to an integer $\Delta n$ known as ``topological charge'',
\begin{align}\label{eq:Deltan}
\left.A_\mu\right|_{\partial\Omega}= {\rm i}{} U(x)\partial_\mu U^{\dagger}(x), \,U(x)\in{\rm SU}(3)\Rightarrow\int_\Omega d^4x \frac{1}{16\pi^2}\, \theta\, {\rm tr}\,F \widetilde F = \theta \Delta n, \, \Delta n\in\mathbb{Z}.
\end{align}
It is known that semiclassical expansions about saddle points of the Euclidean QCD action known as instantons~\cite{Belavin:1975fg,tHooft:1976snw} are sensitive to boundary effects, and hence it is generally expected that nonzero values of the  $\bar\theta$ in Eq.~\eqref{eq:bartheta} will induce nonvanishing predictions for $CP$-violating observables such as the neutron electric dipole moment $d_n$. However, experiments constrain the latter to~\cite{Abel:2020pzs}
\begin{align}
 d_n<1.28\times10^{-26}\,{e\cdot {\rm cm}}.
\end{align}
With traditional theoretical calculations predicting $d_n\propto\bar\theta$, the stringent bounds on $d_n$ are thought to imply a severe tuning of $\bar\theta<10^{-10}$, which is known as the strong $CP$ problem \cite{Callan:1976je,Jackiw:1976pf}.

Here we report on recent calculations that challenge the above conclusion \cite{Ai:2020ptm}, and show a lack of $CP$ violation in the strong interactions with and without relying on instantons methods. For brevity, in these proceedings we will not cover the instanton-based calculations and will focus instead on arguments using  a minimal set of assumptions  related to mathematical consistency in the choice of boundary conditions, { cluster decomposition} \cite{Weinberg:1996kr}, and the index theorem \cite{Atiyah:1963zz}.


\section{The importance of boundary conditions in the path integral}

The Euclidean QCD partition function is meant to correspond to a transition amplitude from the vacuum onto itself after a time interval $T$,
\begin{align}\label{eq:ZQCD}
 Z[T]=\langle 0| {\rm e}^{-H T} |0\rangle={\rm e}^{-E_0 T}.
\end{align}
On the other hand, path integrals correspond to transition amplitudes between eigenstates of the field operators in the Heisenberg picture. Denoting the different fields collectively as $\phi$, one has
\begin{align}\label{eq:pathint}
 {\int_{\phi(-T/2)=\bar\phi}^{\phi(T/2)=\bar\phi'}} {\cal D}\phi {\rm e}^{-{S_E[\phi;\,T]}}= {}_{T/2}\langle\bar\phi'|{\rm e}^{-H T} |\bar\phi\rangle_{-T/2}=\sum_n {}_{T/2}\langle\bar\phi'|n\rangle\langle n|\bar\phi\rangle_{-T/2} {\rm e}^{-E_n T}.
\end{align}
Above, the path integration is such that the field trajectories are assumed to tend to $\bar\phi',\bar\phi$ for $t\rightarrow -T/2,T/2$, respectively. $|\bar\phi\rangle_t$ are eigenstates of the Heisenberg-picture field operators $\hat\phi(t)$, with eigenvalue $\bar\phi$.  Before the last equality in Eq.~\eqref{eq:pathint} we have inserted the spectral decomposition of the identity into eigenstates of the Hamiltonian, $ H|n\rangle = E_n|n\rangle, \sum_n |n\rangle\langle n|=\mathbb{I}$. Clearly, the path integration in Eq.~\eqref{eq:pathint} is a very different object than the vacuum partition function in Eq.~\eqref{eq:ZQCD}. To recover the latter, one can consider two options, as detailed next.

First, one might take $T\rightarrow\infty$, such that for finite $\langle \bar\phi'|n\rangle,\langle n|\bar\phi\rangle$ the contributions of the excited states in Eq.~\eqref{eq:pathint} become infinitely suppressed with respect to the vacuum contribution. With $T\rightarrow\infty$, the spacetime volume $VT$ goes to infinity, and the boundary conditions remain arbitrary. It is generally expected from {} Picard-Lefschetz theory that the full path integral with {open} boundary conditions can be expressed as a sum over  path integrations over complexified  steepest descent field paths, each of them passing through  a saddle point of the Euclidean action \cite{Witten:2010cx}. Over each  path, {$\exp(-S_E)$ is maximal} 
at the saddle, {so that if the saddles have $S_E>0$} one only  needs to consider those with finite action. For $VT\rightarrow\infty$, finite action requires the fields to approach pure gauge configurations asymptotically. In this case the condition in Eq.~\eqref{eq:Deltan} applies, and the saddle points fall into equivalence classes with integer topological charge $\Delta n$. As the steepest descent field paths can be described by means of a continuous deformation from the saddle points, we expect each path to fall into a single equivalence class. Hence for $VT\rightarrow\infty$ one can write 
\begin{align}\label{eq:Zsum}
 Z=\sum_{\Delta n=-\infty}^\infty Z_{\Delta n} \text{ \,\,\,                              for infinite spacetime   }(T\rightarrow\infty).
\end{align}
As the boundary conditions for infinite $T$ are arbitrary, there is nothing wrong a priori with choosing fixed boundary conditions. In particular, one could restrict the path integration to a single topological sector $\Delta m$ at infinite volume. As $\Delta m$ is gauge invariant, so is $Z_{\Delta m}$. In this case, $\theta$ would always enter all correlators through a global phase $\exp({\rm i}\Delta m\theta)$, which cancels { out} when normalizing the correlators by the partition function. Hence there can't be any $CP$-violation in a fixed topological sector. As will be seen, this conclusion still holds for 
$Z$ as in Eq.~\eqref{eq:Zsum}.

The alternative to an infinite spacetime volume is significantly more challenging. In order to ensure that one projects into the vacuum state, one should integrate over the boundary conditions $\bar\phi,\bar\phi'$ in Eq.~\eqref{eq:pathint} with weights given by the vacuum wave functionals $\langle 0|\bar\phi'\rangle$, $\langle 0|\bar\phi\rangle$. Indeed, by appropriately inserting spectral resolutions of the identity in terms of field eigenstates, one has \cite{Plascencia:2015pga}
\begin{align}\label{eq:pathint2}\begin{aligned}
 \langle0|{\rm e}^{-HT}|0\rangle=&\,\int {\cal D}\bar\phi{\cal D}\bar\phi'\langle0|\bar\phi'\rangle_{T/2}
\,\,{}_{T/2}\langle\bar\phi'|{\rm e}^{-H T} |\bar\phi\rangle_{-T/2}\,\,{}_{-T/2}\langle\bar\phi|0\rangle\\
  =&\,\int {\cal D}\bar\phi{\cal D}\bar\phi'\langle0|\bar\phi'\rangle_{T/2}\,\,{}_{-T/2}\langle\bar\phi|0\rangle
  { \int_{\phi(-T/2)=\bar\phi}^{\phi(T/2)=\bar\phi'}} {\cal D}\phi {\rm e}^{-{{S_E[\phi;\, T]}}}.
\end{aligned}\end{align}
The computation { of} all path integrations in Eq.~\eqref{eq:pathint2} is a daunting task, not least because the vacuum wave functionals are unknown. Moreover, with the boundary conditions $\bar\phi,\bar\phi'$ applied at a finite surface, there is no reason that $\bar\phi,\bar\phi'$ correspond to pure gauge configurations, and as such the classification into equivalence classes with integer topological classes does not apply. Still, such an assumption is usually made in the literature. In this spirit,  starting without a $\theta$ term in the action, the said term and the formal sum of Eq.~\eqref{eq:Zsum} could be recovered for finite $\Omega$  by assuming that the vacuum wave-functionals are a linear combination of $\delta$ functionals with support on static pure gauge configurations (i.e. classical vacua). The latter can be classified with integer Chern-Simons numbers $k$ (so that in an appropriate gauge $\Delta n= k(T/2)-k(-T/2)$) and when demanding gauge invariance the ensuing wave-functional corresponds to the usual $\theta$ vacuum in the literature, $|0\rangle=|\theta\rangle=\sum_k \exp({\rm i} k\theta)|k\rangle$. Such a wave functional with no support outside classical vacua has no clear justification, and certainly  goes against the usual behaviour in quantum mechanics.

In conclusion, the only practical way to ensure that one is computing the vacuum partition function relies on taking the infinite spacetime volume limit, more explicitly taking $T\rightarrow\infty$, regardless of whether the spatial dimensions are infinite or compact without boundaries. Only after taking this limit one can assume that the field configurations fall into classes with integer $\Delta n$.

\section{Correlation functions from the index theorem}

The aim of this section is to show how the functional dependence of the partition functions $Z_{\Delta n}$ on $\theta$, the spacetime volume $\Omega$ and the complex fermion masses $\mc_i,\mc_i^*$ is constrained. Knowing the dependence of $Z_{\Delta n}$ on  $\mc_i,\mc_i^*$ allows one to obtain spacetime-averaged fermion correlators---which should be sensitive to constant $CP$-odd phases coming from $\theta,\alpha_i$---from the partition function with open boundary conditions $Z$. This is because  $\mathfrak{m}_i$ and  $\mathfrak{m}_i^*$ can be seen as sources for integrated two-point functions. As the topological charge $\Delta n$ is only enforced to be an integer for an infinite $\Omega$, we will take the limit $\Omega\rightarrow\infty$ before summing over topological sectors. 
 Within a fixed topological sector $\Delta n$, the spacetime averages of the fermionic correlators can be obtained as
\begin{align}\label{eq:correlatorfixed}\begin{aligned}
 \frac{1}{
 \Omega}\int {\rm d}^4 x\,\langle \bar\psi_i P_{\rm R} \psi_i \rangle_{\Delta n}=&\,-\frac{1}{
 \Omega}\frac{\partial}{\partial \mc_i}Z_{\Delta n}, \\
 \frac{1}{
 \Omega}\int {\rm d}^4 x\,\langle \bar\psi_i P_{\rm L} \psi_i \rangle_{\Delta n}=&\,-\frac{1}{
 \Omega}\frac{\partial}{\partial \mc^*_i}Z_{\Delta n},
\end{aligned}\end{align}
as should be clear from Eq.~\eqref{eq:topologcicalterm} and the fact that the path integrals are weighted by $\exp(-\int d^4x {\cal L})$. Let's consider now a partition of  $\Omega$ into subvolumes $\Omega_1$ and $\Omega_2$. Following Ref.~\cite{Weinberg:1996kr}, noting that $\Delta n$ is a surface integral allows to write the partition function as
\begin{align}\label{eq:Zcluster}
 Z_{\Delta n}(\Omega)=\sum_{\Delta n_1=-\infty}^\infty Z_{{\Delta n}_1}(\Omega_1) Z_{{\Delta n}-{\Delta n}_1}(\Omega_2).
\end{align}
Next, we aim to express $Z_{\Delta n}$ as a product of phases times a real function. First, there is the  phase $Z_{\Delta n}(\Omega)\propto {\rm e}^{\ic \Delta n\theta}$ that follows from $Z_{\Delta n}$ being a path integration over $\exp(-{\cal L})$ and from Eqs.~\eqref{eq:topologcicalterm}, \eqref{eq:Deltan}. Other complex phases can only come from the $\alpha_i$ that enter through the fermionic path integration. At one-loop order about each saddle-point, the fermionic path integration leads to a determinant of the massive Dirac operator $\slashed{D}+\mc_i P_R+\mc_i^* P_L$ at the saddle {for each flavour}. The phase of the determinant is fixed by parity and the Atiyah-Singer index theorem~\cite{Atiyah:1963zz}. Indeed, parity relates eigenvalues of eigenfunctions not annihilated by $\slashed{D}$, which come in mutually conjugate pairs and hence do not contribute to the phase \cite{Ai:2020ptm}. The latter is then fully determined by the zero modes of $\slashed{D}$. Given the structure of the operator $\slashed{D}+\mc_i P_R+\mc_i^* P_L$, right/left-handed zero modes have phases ${\rm e}^{\pm\ic \alpha_i\theta}$, respectively, and the total phase per fermion flavour  $i$ is equal to $\alpha_i$ times the difference in the  number of right and left-handed zero modes. But according to the index theorem, this quantity coincides with $\Delta n$. Accounting for all flavours, 
one gets a phase ${\rm e}^{\ic \Delta n\bar\alpha}$. Together with the $\theta$-dependent phase and using Eq.~\eqref{eq:topologcicalterm}, this gives
\begin{align}\label{eq:Ztildeg}
 Z_{\Delta n}(\Omega) =  {\rm e}^{\ic \Delta n\bar\theta}\tilde g_{\Delta n}(\Omega)\quad \text{   with     } \quad\tilde g_{\Delta n}(\Omega)\in\mathbb{R}.
\end{align} Equation~\eqref{eq:Zcluster} then yields the relations
\begin{align}\label{eq:clustertilde0}
 \tilde g_{{\Delta n}}(\Omega_1+\Omega_2)=\!\!\sum_{{\Delta n}_1=-\infty}^\infty\tilde g_{{\Delta n}_1}(\Omega_1) \tilde g_{{\Delta n}-{\Delta n}_1}(\Omega_2).
\end{align}
Setting $\Omega_2=0$ gives  $\tilde g_{\Delta n}(0) = \delta_{\Delta n,0}$. Under parity, $\Delta n$ changes sign, and since the real $\tilde g_{\Delta n}(0)$ are insensitive to $CP$-odd phases, one must have $\tilde g_{-\Delta n}(\Omega)=\tilde g_{\Delta n}(\Omega)$. This motivates  {the  ansatz}
\begin{align}
\label{eq:Ansatz0}
 \tilde g_{\Delta n}(\Omega)=\Omega^{|{\Delta n}|} f_{|{\Delta n}|}(\Omega^2), \quad f_{|{\Delta n}|}(0)\neq0.
\end{align}
Under the assumption of analyticity in $\Omega$, the solution turns out to be unique, giving in the end~\cite{Ai:2020ptm}
\begin{align}\label{eq:Zbeta0}
 Z_{\Delta n}(\Omega) =  I_{\Delta n}(2\beta\Omega)   \,{\rm e}^{\ic \bar\theta\Delta n}\,,
\end{align}
where $I_{\Delta n}$ are modified Bessel functions and $\beta$ is a real parameter that can only depend on the moduli of the $\mathfrak{m}_i$: $\beta=\beta(\mathfrak{m}_j\mathfrak{m}_j^*)$.  From Eqs.~\eqref{eq:correlatorfixed}, \eqref{eq:Zbeta0}, normalizing by the full partition function and summing over $\Delta n$  after taking the limit $\Omega\rightarrow\infty$, we obtain the following correlators,
\begin{align}\label{eq:finalcorrelators}\begin{aligned}
  \frac{1}{\Omega}\int d^4 x\,\langle \bar\psi_i P_{\rm R} \psi_i \rangle\equiv \lim_{N\rightarrow\infty}\lim_{\Omega\rightarrow\infty}\frac{\sum_{|\Delta n|<N}\int d^4x \langle \bar\psi_i P_{\rm R} \psi_i \rangle_{
\Delta n}}{\Omega \sum_{|\Delta m|<N} Z_{\Delta m}} = {-}2\mc^*_i\,\partial_{\mc_i \mc^*_i} \beta(\mc_k \mc^*_k),\\
  \frac{1}{\Omega}\int d^4 x\,\langle \bar\psi_i P_{L} \psi_i \rangle \equiv\lim_{N\rightarrow\infty}\lim_{\Omega\rightarrow\infty}\frac{\sum_{|\Delta n|<N}\int d^4x \langle \bar\psi_i P_{\rm L} \psi_i \rangle_{
\Delta n}}{\Omega \sum_{|\Delta m|<N} Z_{\Delta m}} ={-}2\mc_i\,\partial_{\mc_i \mc^*_i} \beta(\mc_k \mc^*_k).
\end{aligned}\end{align}
To arrive to the previous expressions, we used $\lim_{x\to\infty}I_{\Delta n}(x )/I_{\Delta n^\prime}(x)=1$.
The results remain $\theta$-independent
and the correlators have phases which, being aligned with those of the tree-level masses, can be rotated away with suitable chiral field redefinitions. By taking additional derivatives with respect to the masses $\mathfrak{m}_j$,$\mathfrak{m}_j^*$, the results can be extended to correlation functions involving more fermion fields. While the field redefinitions that yield real correlators  change $\theta$, this has no impact on correlators/physical observables as they are $\theta$-independent. Hence, there is no $CP$ violation.  This conclusion depends crucially on the order of the limits in Eqs.~\eqref{eq:finalcorrelators}. Using the (mathematically incorrect) opposite ordering, one obtains a dependence on both $\theta$ and the $\alpha_i$, so that  observables would be predicted to depend on $\bar\theta$, recovering the traditional picture.

\section{Discussion}

Our results for spacetime-averaged fermion correlators in QCD imply that there cannot be any $CP$-violation coming from nonperturbative corrections to fermionic interactions. 
Both the form of the partition functions in Eq.~\eqref{eq:Zbeta0} and the alignment of correlators with the phases $\alpha_i$ can be recovered as well with instanton calculus \cite{Ai:2020ptm}.  When matching our results for fermion correlators with effective operators, in particular the 't~Hooft interactions with $2N_f$ fermion fields that are expected from anomalies, one infers that the phases of the `t~Hooft vertices are fixed by the $\alpha_i$:
\begin{align}
\label{eq:eff:operator}
{\cal L}_{\rm eff}\supset-\Gamma_{N_f} {\rm e}^{{\rm i}\xi} \prod_{j=1}^{N_f}(\bar\psi_j P_{\rm L}\psi_j)-\Gamma_{N_f}{\rm e}^{-{\rm i}\xi}\prod_{j=1}^{N_f}(\bar\psi_j P_{\rm R}\psi_j)\,, \quad \xi=-\bar\alpha.
\end{align}
The former result applies regardless of whether one sums over topological sectors in the partition function or not. The only other QCD computation of the phase in the `t~Hooft vertices that we are aware of is that of `t~Hooft himself, who used instanton calculus in the dilute gas approximation, but assumed the opposite order of limits {than the one } used in Eq.~\eqref{eq:finalcorrelators} and concluded that the phase $\xi$ in Eq.~\eqref{eq:eff:operator} was $\xi=\theta$ \cite{tHooft:1976snw,tHooft:1986ooh}. Again, mathematical consistency demands the ordering of limits used in eq.~\eqref{eq:finalcorrelators}, leading to $\xi=-\bar\alpha$ (which complies with chiral symmetries) and no $CP$ violation.

The effective interactions in Eq.~\eqref{eq:eff:operator} can be matched to the determinant terms in the chiral Lagrangian, whose phases are then predicted to be fixed by $\bar\alpha$. This leads to no $CP$-violation in the low-energy effective theory, and in particular no dipole moment for the neutron, while still leading to an enhancement of the $\eta'$ mass \cite{Ai:2020ptm}.

\end{document}